\documentclass[5p,number,sort&compress]{elsarticle} 
\usepackage{graphicx} 
\usepackage{color}	
\usepackage[]{hyperref}
\hypersetup{
 colorlinks=true,
 linkcolor={blue},
 citecolor={blue},
 urlcolor={blue},
 linktocpage=true,
 breaklinks=true,
}

\usepackage{lineno}
\modulolinenumbers[5]

\begin{document}


\newcommand{\LBG}[1]{\textcolor{blue}{[LBG: #1]}} 
\newcommand{\AL}[1]{\textcolor{red}{[AL: #1]}} 
\newcommand{\DR}[1]{\textcolor{orange}{[DR: #1]}} 
\newcommand{\PM}[1]{\textcolor{magenta}{[PM: #1]}} 
\newcommand{\MR}[1]{\textcolor{purple}{[MR: #1]}} 
\title{
Performance of the Particle-Identification Silicon-Telescope Array Coupled with the VAMOS++ Magnetic Spectrometer
}

\author[1,2]{L.~Bégué--Guillou}
\author[1]{A.~Lemasson\corref{cor1}}
\author[2,3]{P.~Morfouace}
\author[1]{D.~Ramos}
\author[2,3]{J.~Taieb}
\author[1]{J.D.~Frankland}
\author[1]{M.~Rejmund}
\author[1]{G.~Fremont}
\author[1]{P.~Gangnant}
\author[1]{A.~Cobo-Zarzuelo}
\author[1]{N.~Kumar}
\author[2,3]{T.~Efremov}
\author[2,3]{A.~Chatillon}
\author[1]{E.~Clément}
\author[1]{G.~De~France}
\author[1]{A.~Francheteau}
\author[1]{I.~Jangid}
\author[2,3]{C.~Lenain}
\author[2,3]{B.~Mauss}
\author[1]{T.~Tanaka}
\author[7]{L.~Audoin}
\author[4]{M.~Caamano}
\author[4]{B.~Errandonea}
\author[7]{M.~Godio}
\author[6]{D.~Gruyer}
\author[1]{B.~Jacquot}
\author[1]{M.~Lalande}
\author[8]{R.~C.~Malone}
\author[4]{A.~Munoz}
\author[9]{A.~P.~D.~Ramirez}
\author[4]{J.~L.~Rodr\'{i}guez-S\'{a}nchez}
\author[5]{C.~Schmitt}
\author[2,3]{O.~Syrett}
\author[2,3]{C.~Surrault}
\author[9]{A.~P.~Tonchev}

\cortext[cor1]{%
\begin{minipage}[t]{\textwidth}
\hspace*{0.1cm}Corresponding author\\
\hspace*{0.1cm}\textit{E-mail address} \href{mailto:antoine.lemasson@ganil.fr}{antoine.lemasson@ganil.fr} (A. Lemasson)
\end{minipage}}


\affiliation[1]{organization={GANIL, CEA/DRF-CNRS/IN2P3},
postcode={F-14076},
city={Caen},
country={France}}

\affiliation[2]{organization={CEA, DAM, DIF},
postcode={F-91297},
city={Arpajon},
country={France}}

\affiliation[3]{organization={Université Paris-Saclay, CEA, LMCE},
postcode={F-91680},
city={Bruyères-le-Châtels},
country={France}}

\affiliation[4]{organization={IGFAE -- Universidade de Santiago de Compostela},
postcode={E-15706},
city={Santiago de Compostela},
country={Spain}}

\affiliation[5]{organization={Institut Pluridisciplinaire Hubert Curien, CNRS/IN2P3-UDS},
postcode={F-67037},
city={Strasbourg},
country={France}}

\affiliation[6]{organization={LPC Caen, ENSICAEN, Université de Caen, CNRS/IN2P3},
postcode={F-14000},
city={Caen},
country={France}}

\affiliation[7]{organization={IJCLab,Université Paris-Saclay, CNRS/IN2P3},
postcode={F-91405},
city={Orsay},
country={France}}

\affiliation[8]{
organization={Volgenau Physics Department, United States Naval Academy},
postcode={MD 21409},
city={Annapolis},
country={USA}
}

\affiliation[9]{organization={Nuclear and Chemical Sciences Division, LLNL},
postcode={CA, 94550},
city={Livermore},
country={USA}}


\begin{abstract}

The Particle-Identification Silicon-Telescope Array (PISTA) is a new detection system designed for high-resolution studies of fission process
induced by multi-nucleon transfer in inverse kinematics. It is specifically optimized for experiments with the {VAMOS++} magnetic spectrometer at GANIL (Grand Accélérateur National d’Ions Lourds). The array comprises 
eight trapezoidal $\Delta$E-E silicon telescopes arranged in a lamp shade configuration. Each telescope integrates two single-sided 
stripped silicon detectors, enabling target-like recoil identification, energy loss measurements, and trajectory reconstruction. 
Positioned in close proximity to the target, PISTA’s compact geometry achieves high-efficiency tracking of target-like recoils 
produced in multi-nucleon transfer reactions at Coulomb barrier energies. The spatial segmentation of the array allows precise determination 
of the mass and charge of the target-like nucleus, and excitation energy of fissioning systems. 
This work presents the particle identification and excitation energy reconstruction performances for the interactions of $^{238}$U 
beam with $^{12}$C target. An excitation energy resolution of $800$~keV (FWHM) was determined together with mass resolution of $1.1$~\% (FWHM). 
The combination of PISTA and {VAMOS++} magnetic spectrometer enables unprecedented investigations of the fission process 
as a function of the excitation energy of the fissioning nucleus, particularly for exotic systems produced in transfer-induced reactions. 

\end{abstract}

\sloppy

\maketitle
\section{Introduction}
A thorough comprehension of the fission process necessitates an in-depth understanding of the influence of nuclear properties and 
the characteristics of the reaction entrance channel on the evolution of the fissioning system from its initial configuration to the formation of fission fragments. The excitation energy ($E^{\ast}$) of the fissioning system significantly influences both the probability of fission and the characteristics of the resulting fragments. Measurements of fission observables as a function of 
$E^{\ast}$ provide crucial insights into the interplay between nuclear structure effects, dissipation, and the dynamics of 
the large-amplitude deformation process leading to scission~\cite{Andreyev2018}. The evolution of fission fragment yields and fission probabilities with respect to excitation  energy is significant for both fundamental nuclear physics investigations, encompassing the study of shell effects and their damping as excitation increases, and practical applications within the nuclear industry, such as reactor design and waste transmutation.

Systematic studies of fission probabilities~\cite{Perez2020,Kean2019,rodriguez2014} and fragment yields~\cite{Hirose2017,Vermeulen2020,Ramos2019,Ramos2023} as function of $E^{\ast}$ for various compound nuclei across the nuclear chart are fundamental to constrain theory, including model ingredients such as level densities, barrier
heights, and the role of shell effects. Provided that the fissioning system and its excitation energy can be measured event-by-event, multi-nucleon transfer induced reactions are ideal to that goal, since several systems can be studied in a single experiment~\cite{Caamano2013,rodriguez2014,Ramos2019,Ramos2020,Kean2019,Vermeulen2020}.

Traditionally, isotopic yields have been measured using thermal neutron-induced fission reactions in setups such as the Lohengrin spectrometer \cite{LANG198034, SCHMITT198421}. However, such methods often suffer from limited resolution in fission fragment charge identification, restricting access to isotopic yields only for the light fission fragment. The challenges are even greater with non-thermal neutron beams, due to a significant reduction in the available neutron flux and a decreased fission rates. Recent experimental programs employ fast neutron beams \cite{MEIERBACHTOL201559, Falstaff2025} to address these challenges by measuring fission fragment yields as 
a function of neutron energy. Alternatively, cumulative fission fragment yields have been measured via activation technique combined with high-resolution $\gamma$-ray spectroscopy following irradiation at different incident neutron energies~\cite{gooden2016, gooden2024, tonchev2025}.

In the last decades, elemental and later isotopic yields with high resolution have been measured using Coulomb-excitation reactions 
in inverse kinematics with fast radioactive beams at GSI  using the FRS spectrometer and SOFIA setup~\cite{schmidt2000, chatillon2020, pellereau2017accurate}. While  this method provides unique access to a wide range of systems, the excitation energy of the fissioning nucleus cannot be determined  on an event-by-event basis. Alternatively, in multi-nucleon transfer reactions at energies around the Coulomb barrier, the fissioning system is produced as a heavy partner, and its properties such as mass, atomic number, and excitation energy can be inferred from the measurement of the complementary, light partner. In Refs~\cite{Leguillon2016,Hirose2017,Kean2019,Vermeulen2020}, multi-nucleon transfer in 
a direct-kinematics framework were employed to identify the fissioning system and to extract its excitation energy. However, in direct 
kinematics, complete isotopic identification of the fission fragments cannot be achieved due to the very limited velocity of the fission fragments.

The combination of inverse kinematics, and multi-nucleon transfer reactions enables the simultaneous measurement of entrance channel properties and isotopic identification of fission fragments~\cite{Caamano2013,rodriguez2014}. This approach was initially implemented at the {VAMOS++} spectrometer at 
GANIL \cite{Rejmund2011} using the SPIDER silicon telescope array where isotopic yields and fission probabilities were reported in Refs.~\cite{Caamano2013,rodriguez2014,Ramos2019,Ramos2020}.
However the fissioning system excitation energy resolution and target-like recoil isotopic separation were constrained by the design of the target-like recoil detection system, i.e SPIDER array, specifically its insufficient spatial granularity.

To address these limitations, a new detection system, PISTA (Particle Identification Silicon Telescope Array), was developed for the measurement and characterization of target-like 
recoils from multi-nucleon transfer reactions. PISTA enables isotopic identification of the coincident reaction partner nucleus 
(which may undergo fission) and a precise determination of its excitation energy. The implementation with the {VAMOS++} spectrometer, 
which provides isotopic identification of fission fragments, allows the correlation of the entrance and exit channels of the reaction on 
an event-by-event basis. This capability allows for more precise studies of fission probabilities and fission fragment yields as a function of $E^{\ast}$ for a wide range of systems.

This paper presents a characterization of the PISTA device, detailing its design, construction, and performances. Notably, it highlights 
the device’s capability of achieving high-precision isotopic separation, accurate excitation-energy reconstruction, and its suitability for operation in inverse kinematic configurations, which are particularly advantageous for studies of the fission process.

\begin{figure}[h!]
\begin{center}
\includegraphics[width=\columnwidth]{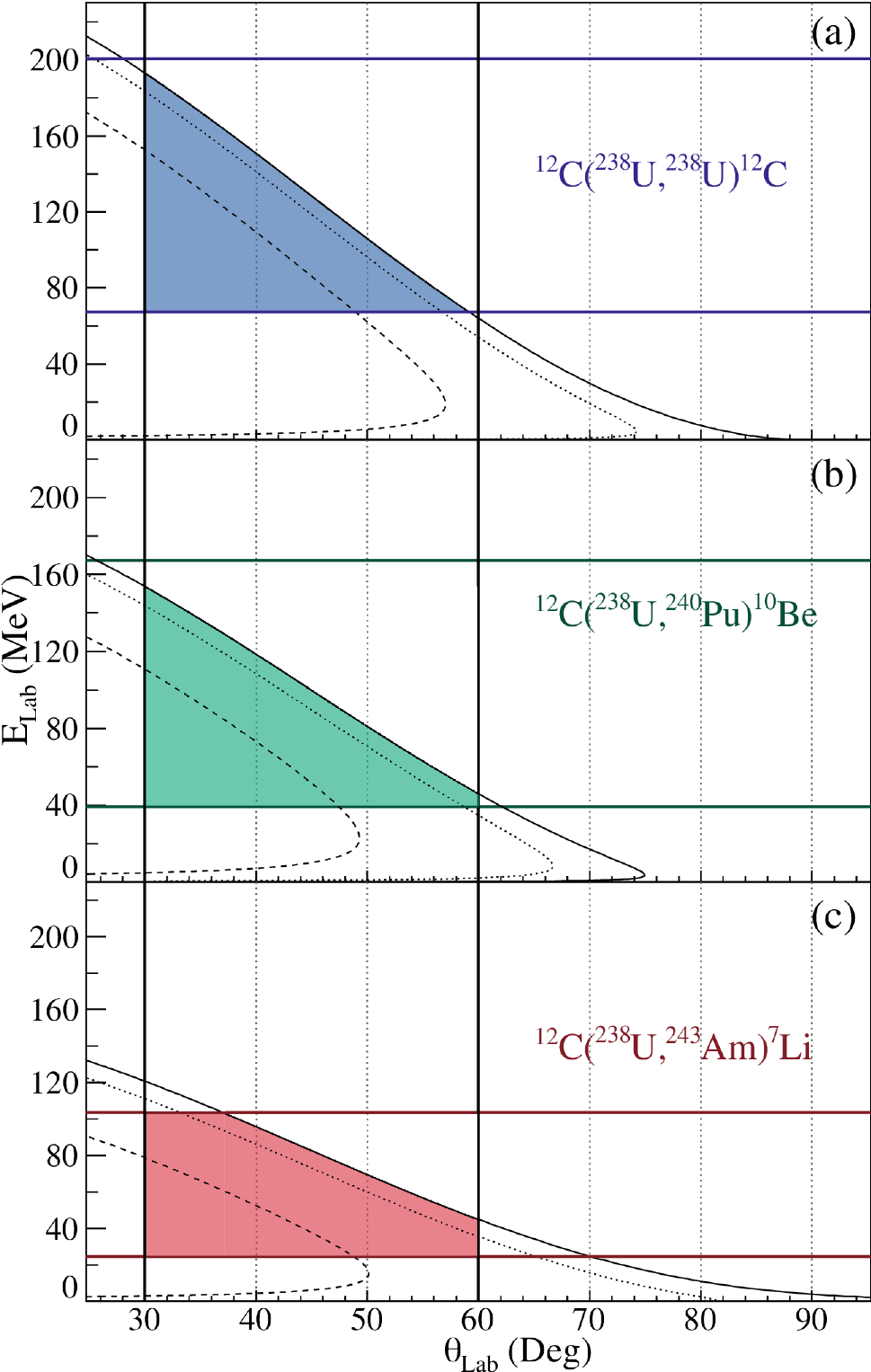}
\end{center}
\caption{\label{pic:kinelines}
Calculations for the reactions in inverse kinematics
(a) $^{12}$C($^{238}$U,$^{238}$U)$^{12}$C, 
(b) $^{12}$C($^{238}$U,$^{240}$Pu)$^{10}$Be, and 
(c) $^{12}$C($^{238}$U,$^{243}$Am)$^{7}$Li, 
where a $^{238}$U beam at 5.95 MeV/A is bombarding a $^{12}$C target.
The distinct curves illustrate the laboratory energy-angle correlation of the recoiling target-like nucleus for reactions resulting 
in the ground state (solid line) and excited states at $E^{\ast}=5$~MeV (dotted line) and $E^{\ast}=20$~MeV (dashed line) in the beam-like heavy ejectile. 
The vertical lines denote the angular range encompassed by PISTA, while the horizontal lines correspond to the energy constraint necessary 
for complete identification (i.e., events where the target-like ions are stopped in the second stage of the silicon telescope).
 }
\end{figure}
\section{Detection requirements}

The primary objective of the described setup is to use two-stage silicon telescopes and apply the $\Delta$E-E technique to isotopically identify 
target-like nuclei spanning the range from helium to oxygen produced in multi-nucleon reactions in inverse kinematics. Heavier nuclei can be also identified provided that they can penetrate the $\Delta$E stage with sufficient energy. The excitation energy of the fissioning system is inferred using the missing-mass method~\cite{Suzuki2012}. Accurate application of the missing-mass 
method  requires precise determination of the scattering angle ($\theta_{lab}$) and the kinetic energy ($E_{lab}$) of the target-like residue. 

Figure~\ref{pic:kinelines} illustrates the calculated typical kinematical correlation between the kinetic energy of a target-like residue and its 
laboratory angle. The reactions are induced by a $^{238}$U beam at a bombarding energy of $5.95$~MeV/A on a $^{12}$C target. 
Panels (a)-(c) represent three  distinct reaction channels of interest, as follows:
(a) $^{12}$C($^{238}$U,$^{238}$U)$^{12}$C,
(b) $^{12}$C($^{238}$U,$^{240}$Pu)$^{10}$Be, and
(c) $^{12}$C($^{238}$U,$^{243}$Am)$^{7}$Li. 
The distinct curves illustrate the laboratory energy-angle correlation of the target-like nucleus for reactions resulting in various 
excitation energy ($E^{\ast}$) within the beam-like ejectile.

To fully identify target-like recoils in a two-stage silicon telescope and reconstruct their kinetic energies, ions must possess sufficient energy 
to traverse the first stage ($\Delta E$), while remaining low enough in energy to stop within the second stage ($E$). This requirement, 
intimately linked to the thicknesses of the two stages, defines the corresponding accessible energy intervals as illustrated by the horizontal
lines in Figure~\ref{pic:kinelines}.  A critical design requirement was to prevent fission fragments from interacting with the silicon detector 
array, which would both hinder identification  and pose a risk of damage to the silicon detectors. Fission fragments emitted in-flight for 
all reactions of interest are confined to a cone of less than $30^\circ$, which defines the inner angular limits of the detector array. 
The vertical lines in Figure~\ref{pic:kinelines} denote the angular range encompassed by PISTA, while 
the shaded regions indicate the angular and energy phase space accessible for each reaction.

A high-resolution isotopic identification necessitates a uniform effective thickness across a broad angular range in the first detection stage  ($\Delta E$). In previous studies employing the SPIDER array \cite{rodriguez2014, Ramos2020, Ramos2019}, the detector was positioned  perpendicular to the beam axis, resulting in a significant variation in effective thickness as a function of the polar angle of reaction products.  This variation posed a particular challenge along with large strip sizes. Consequently, segmented detectors with a thickness uniformity better than $1$~\% and a detector surface oriented nearly perpendicular to the particle trajectory are essential to minimize the dependence of the effective detector thickness on the particle’s incident angle.

Furthermore, PISTA aims to enhance the excitation-energy resolution compared to previous studies conducted using the SPIDER annular silicon detection system. In these studies, an excitation-energy resolution of $2.7$~MeV (FWHM) was reported \cite{rodriguez2014, Ramos2020, Ramos2019}.  The excitation-energy resolution of the fissioning system is primarily determined by the accuracy of the scattering-angle determination and  the target thickness. The beam spot size, typically $\sigma_x \sim 0.6$ mm and $\sigma_y \sim 1 $ mm, exceeds the intended detector strip size of $0.535$~mm. Consequently, determining the scattering angle necessitates both the interaction position in PISTA detector and the  beam interaction position on the target. The beam interaction position can be determined using the Dual Position Sensitive Multi-Wire Proportional Counter (DPS–MWPC)  tracking detector used with {VAMOS++} \cite{Vandebrouck2016}.  To achieve an excitation-energy resolution better than $1$~MeV, the required angular resolution must not exceed $0.18^\circ$,  and the energy resolution must not exceed $800$~keV (including uncertainty on the reaction position in the target). Furthermore, a substantial solid angle should be encompassed to compensate for the relatively small cross sections of the channels of interest.  

To fulfill these requirements, we implemented an array of single-sided silicon-strip $\Delta E$—$E$ telescopes in a lamp-shade configuration,  which is detailed in the subsequent section.
\begin{figure}[t!]
\centering
\includegraphics[width=\columnwidth]{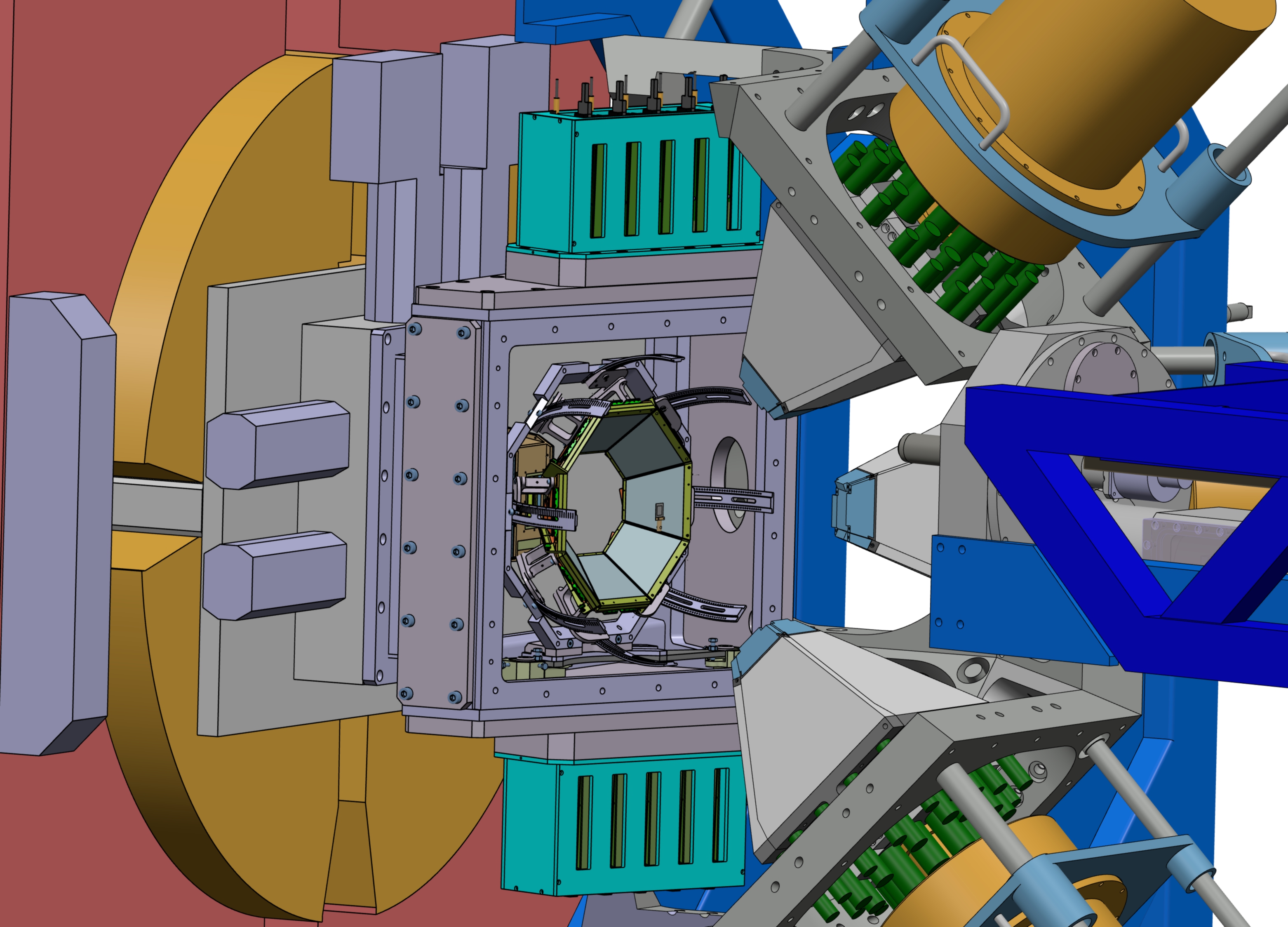}
\caption{\label{pic:target}
Three-dimensional representation of the PISTA detection system is depicted, positioned around the target location in front of the {{VAMOS++}} 
spectrometer~\cite{Rejmund2011}. Three {EXOGAM} HPGe clover detectors~\cite{EXOGAM} are placed at backward angles.
}
\end{figure}

\section{PISTA telescopes}

The PISTA detection system comprises eight trapezoidal telescopes arranged in a lamp-shade geometry, as illustrated in Figure~\ref{pic:target}.
The figure presents a three-dimensional perspective of the PISTA array positioned around the target location in front of the {VAMOS++} spectrometer. 
PISTA's telescopes comprise two silicon stages separated by a $4.5$~mm gap. Furthermore, the setup includes three {EXOGAM} HPGe 
clover detectors~\cite{EXOGAM} placed at backward angles.

\begin{figure}[!t]
\centering
\includegraphics[width=0.9\columnwidth]{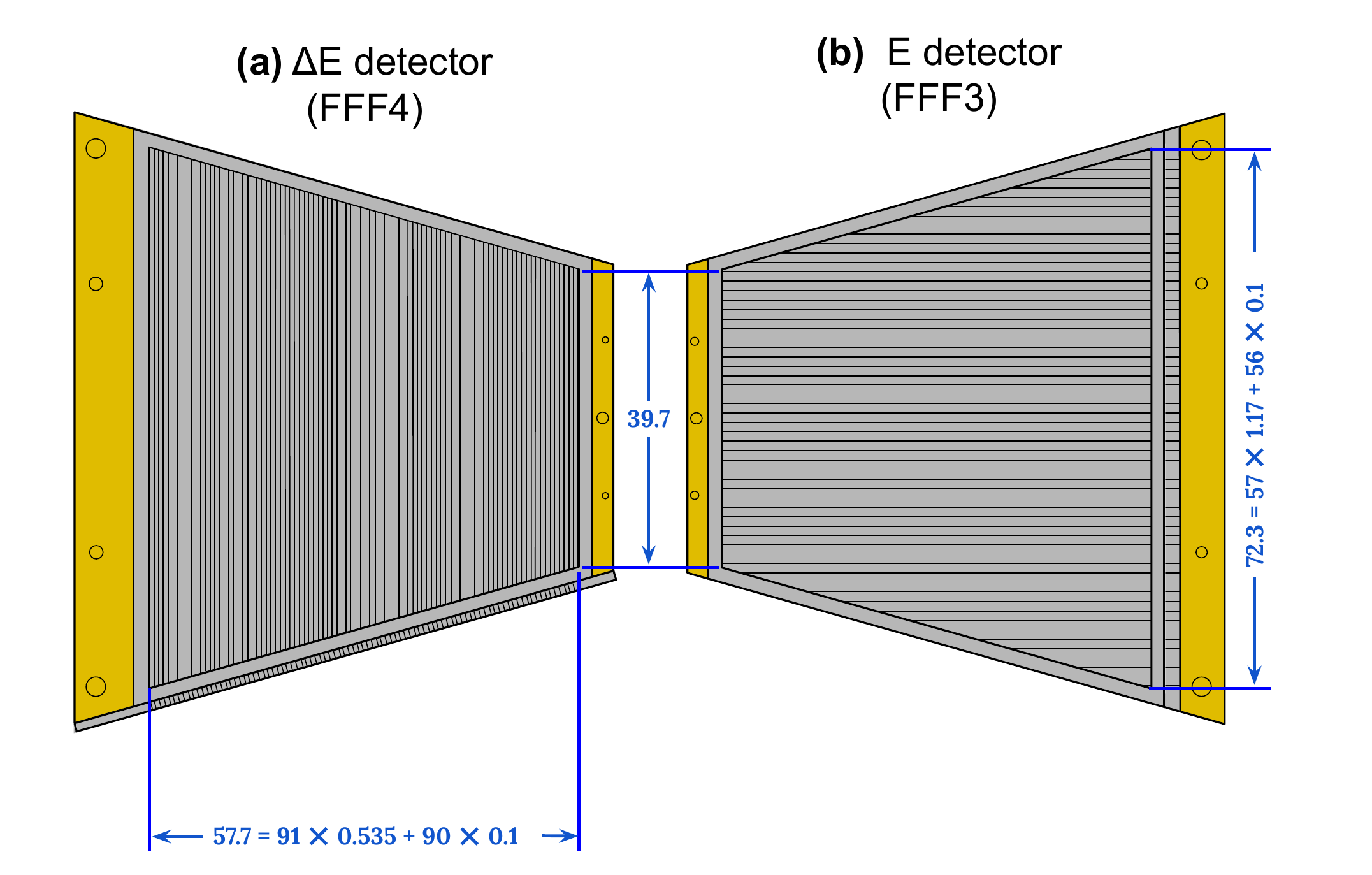}
\caption{\label{pic:pista} 
Drawing of the front side of (a) $\Delta E$ (FFF4 design) and (b) $E$ (FFF3 design) manufactured by Micron Semiconductor Ltd.
}
\end{figure}

\subsection{First stage: $\Delta E$ single-sided detector}

The $\Delta E$ detectors were manufactured by Micron Semiconductor Ltd. using n-type silicon and mounted on custom circuit boards 
designed at GANIL. Figure~\ref{pic:pista}(a) depicts a schematic representation of the front segmented side of the $\Delta E$ stage. 
The detectors have a trapezoidal shape, with their short and long bases measuring $39.74$~mm and $72.29$~mm, respectively, resulting 
in an active area of approximately $27.5$~cm$^2$. Depending on the telescope configuration, the detectors possess thicknesses ranging 
between $100~\mu$m and $108~\mu$m, with a uniformity better than $1\%$.
The entrance side of the detectors is segmented into $91$ strips (as illustrated in Figure~\ref{pic:pista}). The $535~\mu$m wide strips 
are separated by $100~\mu$m passive inter-strips. Due to the geometry, the strips exhibit variable dimensions, leading to variable 
capacitance values ranging between $22.3$ and $40.1$~pF.
It is noteworthy that particular attention was given to the bounding and routing of the strips, which are fabricated on the detector’s side, 
to minimize the loss of solid angle. The rear side is single-readout and serves to polarize the junction with a nominal bias of $20$~V. 
The entrance and exit windows consist of $0.5~\mu$m aluminum, selected to minimize the energy loss of particles within this dead layer. 
The intrinsic energy resolution of the strips and the rear side will be presented in Section~\ref{sec:res}.

\subsection{Second stage: $E$ single-sided detector}

The $E$ detectors were manufactured by Micron Semiconductor Ltd.  using n-type silicon and mounted them on custom circuit boards designed at GANIL. Figure~\ref{pic:pista}(b) depicts a schematic view of the front segmented side of the $E$ stage. The detectors share the same trapezoidal 
shape and dimensions as the $\Delta E$ detectors, but with a slightly larger active area of approximately $29.9$~cm$^2$. The second stage of the PISTA 
telescopes has a typical thickness of $1$~mm. The entrance side of the detectors is segmented into $57$  strips (as shown in Figure~\ref{pic:pista}). 
The $1.17$~mm wide strips are separated by $100~\mu$m passive inter-strips. Due to the geometry, the strips have variable lengths, resulting 
in variable capacitance ranging between $0.2$ and $7.0$~pF. The rear side is single readout and is utilized to polarize the junction with a nominal 
bias of $190$~V. Similar to $\Delta E$ detectors, an aluminum entrance window of $0.5~\mu$m was employed. The intrinsic energy resolution of 
the strips and rear side will be presented in Section~\ref{sec:res}.

\begin{figure}[!t]
\includegraphics[width=\columnwidth]{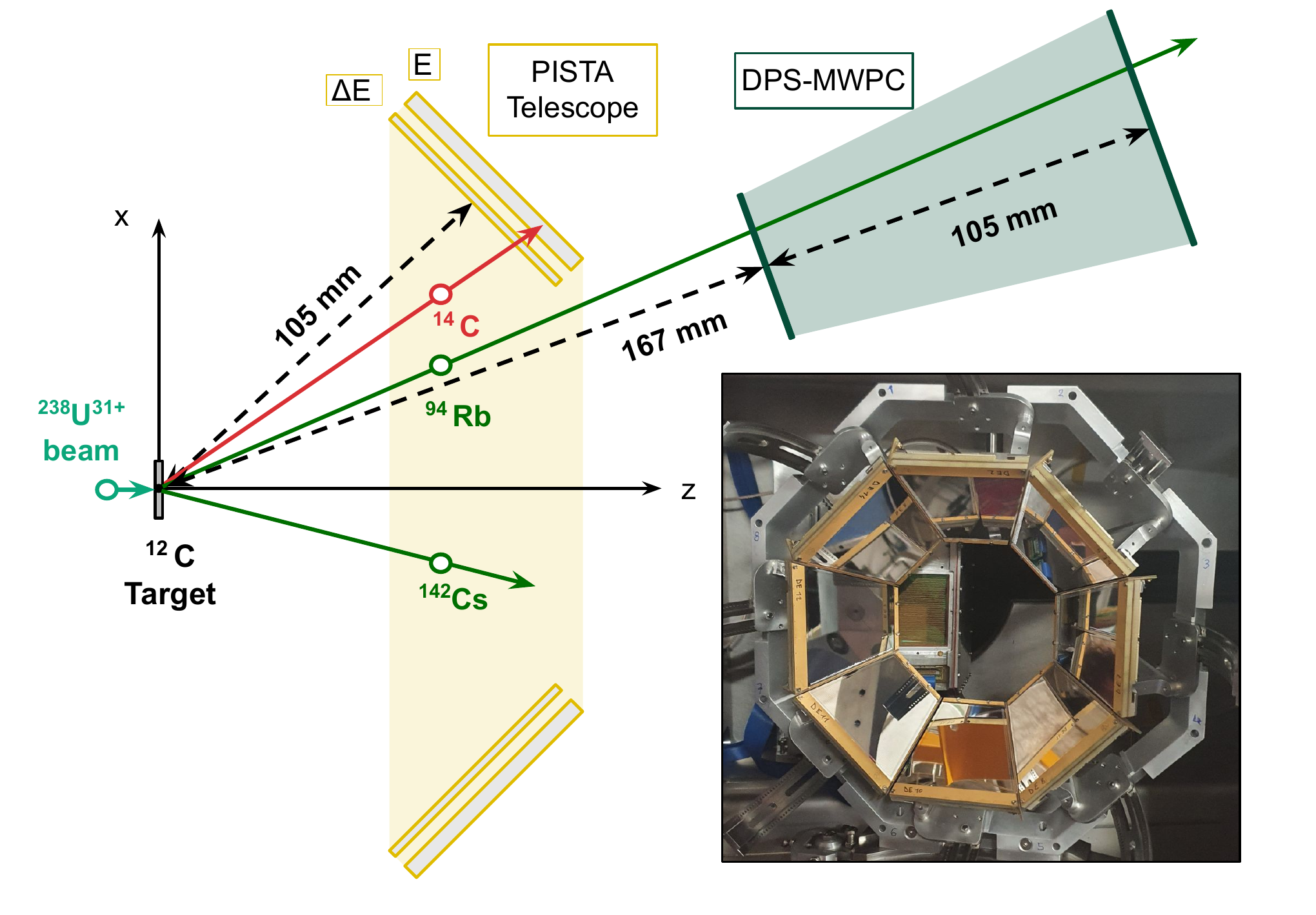}
\caption{\label{pic:pista_mounting}
Schematic top view of the detection systems surrounding the $^{12}$C target. The PISTA array, comprising $\Delta E-E$ telescopes (illustrated in yellow), 
and the DPS-MWPC (illustrated in green) are depicted. A representative reaction scenario is presented for the $^{12}$C($^{238}$U,$^{236}$U)$^{14}$C 
transfer reaction, followed by the fission of $^{236}$U into two fragments, specifically $^{94}$Rb and $^{142}$Cs, and the detection of $^{14}$C in PISTA. 
The inset provides a photograph of the PISTA experimental setup. 
}
\end{figure}

\subsection{Mechanical assembly}

Figure~\ref{pic:pista_mounting} illustrates a schematic implantation of the PISTA array and the DPS–MWPC in front of the {VAMOS++} spectrometer, 
with the relative distances to the $^{12}$C target indicated. The telescopes are mounted on a dedicated holding structure designed to facilitate 
versatile, reproducible, and precise positioning of the telescopes. In the nominal configuration, the detectors form a closed array around the beam axis, 
with each telescope entrance window centered $105.5$~mm away from the target. Each telescope is oriented such that the normal to its center subtends 
a polar angle of $45^\circ$ relative to the beam axis at the target position. In this configuration, the PISTA array covers laboratory polar angles ranging from $30^\circ$ to $60^\circ$. Given the positioning and segmentation of the detector, the expected uncertainties for the target-like trajectory angles are 
approximately $\sigma(\theta_{lab}) \approx 0.12^\circ$ for the polar angle and $\sigma(\phi_{lab}) \approx 0.3^\circ$ for the azimuthal angle. 
This configuration provides suitable angular coverage for target-like recoils while allowing fission fragments to pass through the central aperture up to angles 
of $\theta \sim 30^\circ$. Furthermore, to enable future versatile use of the array for a wider range of kinematics, each telescope can be moved individually 
by steps of $1^\circ$ from $45^\circ$ to $90^\circ$. This modification can be further combined with a change in the installation radius, which is adjusted by utilizing different spacer-mounting components corresponding to predefined radii of $105.5$~mm, $125.5$~mm, and $145.5$~mm.  The target is generally surrounded by permanent magnets (with a magnetic field of $B = 50$~mT), which deflect $\delta$ electrons, produced in the interaction of the highly charged $^{238}$U beam with the target, thereby preventing these electrons from reaching the telescope.

\begin{figure*}[!th]
\includegraphics[width=\textwidth]{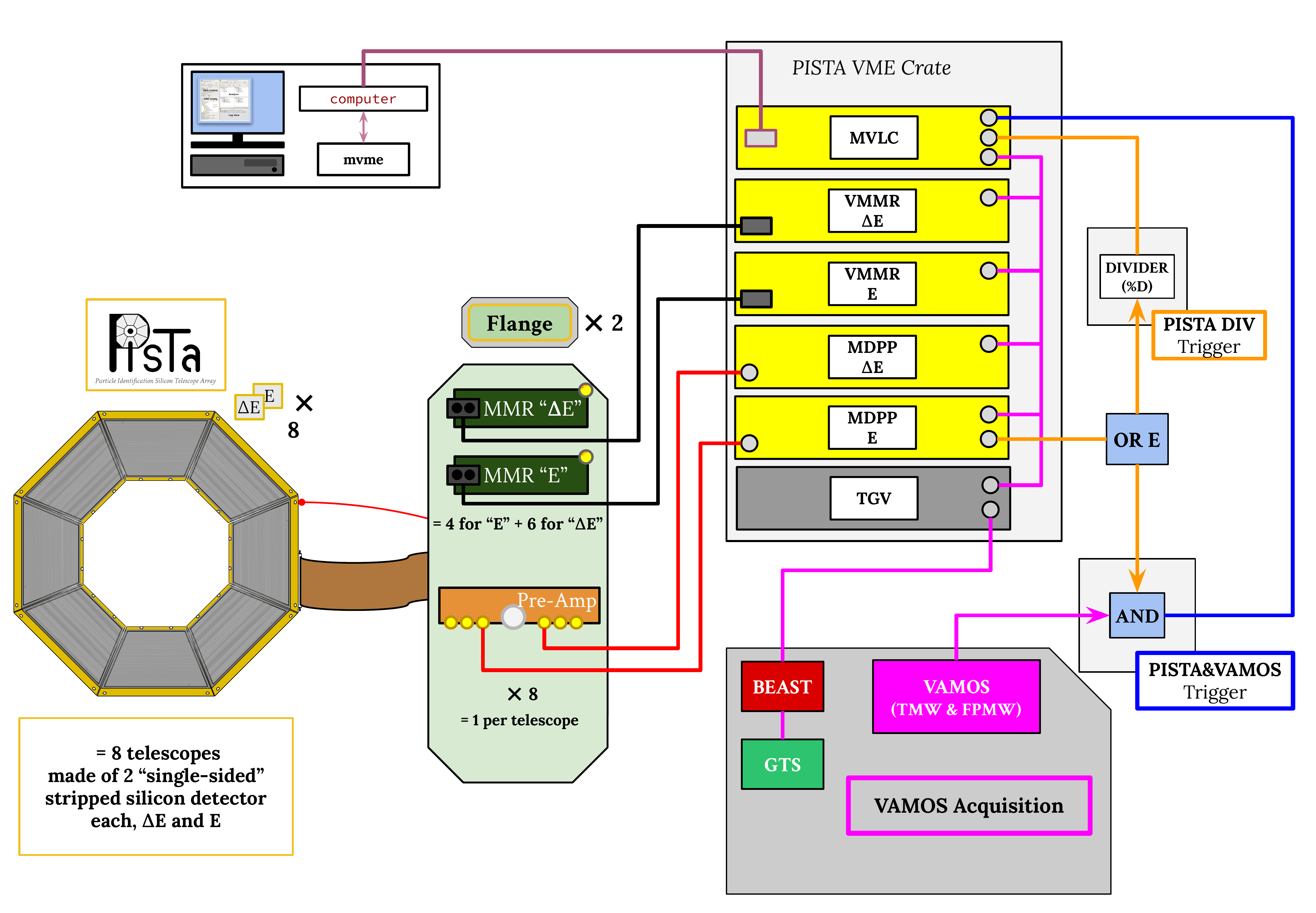}
\caption{\label{pic:Electronics}
Overview of the PISTA hardware connections. On the left, a diagram illustrates the detector array, showing a single $E$ detector 
connected and for half of the front side. The front-side readout is achieved via a flat cable (brown), while the rear side is connected 
using a single coaxial cable (red), which also serves as the bias voltage input. A schematic representation of one electronic flange 
is provided, including a $\Delta E$ MMR and a $E$ MMR module and a preamplifier (Pre-Amp). Each module is annotated with typical 
operating values. On the right, the VME crate dedicated to signal processing is shown, containing all essential modules used for the 
acquisition and digitization of the detector signals. Additionally, in the bottom right corner, the crates responsible for generating 
the trigger logic are displayed, along with the associated electronic modules.}
\end{figure*}
\section{Electronics and data acquisition}
The Data Acquisition (DAQ) system, detector signal digitization, and event building are managed by commercial modules from 
Mesytec GmbH \& Co. The entire electronics setup is housed within a single VME crate, following an implementation similar to 
that reported in Ref.~\cite{Frankland2025}. A comprehensive overview of the hardware connections and PISTA architecture is 
provided in Figure \ref{pic:Electronics}.

\subsection{Readout electronics}
Charge signals from both the front (strips) and rear sides of each telescope detector are processed and recorded through 
a dedicated chain of electronic modules.

\subsubsection{Rear-side detectors\label{sec:elec:rear}}
The 16 rear-side signals ($8~\Delta E$ and $8~E$) are routed to custom charge-sensitive preamplifiers (CSP) developed by CEA-DAM. 
These CSPs were specifically designed for the plain rear-side electrodes of the PISTA detectors.
\begin{itemize}
    \item $\Delta E$ detectors: the 100 $\mu$m $\Delta E$ detectors exhibit a high output capacitance ($>4$~nF), which typically 
    degrades energy resolution. To mitigate this, the $\Delta E$ CSP utilizes three JFETs in parallel to achieve a high input transconductance 
    of $120$~mS, significantly higher than standard commercial units. A high-bandwidth decompensated operational amplifier in 
    the cascade stage ensures a closed-loop bandwidth of $3$~MHz,
    \item $E$ detectors: these detectors have a standard capacitance of approximately $400$~pF. A similar CSP topology is used but with 
    a single JFET, resulting in a transconductance of $45$~mS and a closed-loop bandwidth of $10$~MHz.
\end{itemize}

Digital signal processing and pulse amplitude encoding are performed by two MDPP-16 SCP modules~\cite{MDPP}, one for each detector type.
Within the MDPP, signals are split into two branches:
\begin{itemize}
\item amplitude branch: uses a trapezoidal filter (shaping time $1.2~\mu$s for both $\Delta E$ and $E$) to capture and hold the signal peak,
\item trigger branch: extracts timing information via a Constant Fraction Discriminator (CFD); for $\Delta E$, the shaping time is $1.2~\mu$s and 
for $E$, it is reduced to $0.2~\mu$s.
\end{itemize}

The filters are configured without a flat top, resulting in a triangular output. Signals are only processed if the CFD output falls within 
a "window of interest" (width and offset adjustable) relative to a reference trigger. While these modules can operate in self-triggering 
mode, PISTA utilizes an external validation signal.

\subsubsection{Front stripped side detectors\label{sec:elec:front}}
The $1184$ strip signals are routed to ten Mesytec Multiplexed Readout (MMR) modules~\cite{MMR} (four for $E$, six for $\Delta E$). Each MMR handles up to $128$ channels and performs pre-amplification, digitization, and trigger generation.

The integrated Charge Sensitive Amplifiers (CSA) are ranged for maximum expected energy depositions: $2.7$~pC~($60$~MeV) for 
$\Delta E$ and $9$~pC~($200$~MeV) for $E$. Following the CSA, the signal is split:
\begin{itemize}
\item energy branch: shaped at $500$~ns and passed to a stretcher,

\item trigger branch: shaped at $160$~ns and sent to a leading-edge discriminator.
\end{itemize}

The $128$ amplitude signals are multiplexed across four $50$~MHz ADCs ($4096$ channels). An internal FPGA buffers the digitized 
data and transmits them to VMMR-8 VME modules~\cite{MMR}, which handle de-multiplexing and data formatting. Similar to the MDPP 
modules, the VMMR validates signal readout based on a specific window of interest.

\subsection{Data acquisition trigger}
The VME crate is controlled by an MVLC module~\cite{MVLC}, an FPGA-based controller designed for high-rate data readout. 
The primary PISTA trigger is defined as a logical OR of the rear $E$ detector CFD outputs from the MDPP.

For coincidence experiments, a global trigger is constructed from the logical OR of several sources:

\begin{itemize}
\item PISTA standalone,

 \item PISTA and {VAMOS++} coincidence (generated from the DPS-MWPC detector and the PISTA $E$ signals) using a $100$~ns window.
\end{itemize}

To manage high rates, the PISTA trigger can be downscaled by a factor $D$. Triggers are validated only if the system is not in 
"dead time." Total dead time includes signal processing ($\sim 7-8~\mu$s) and readout ($\sim 6-7\mu$s). Validated triggers are 
distributed by the MVLC to all modules to open their respective windows of interest (typically $1~\mu$s for MMR and $0.5~\mu$s for MDPP).

\subsection{Data acquisition and coupling with other data acquisition systems}

PISTA is integrated into the GANIL ecosystem via the NARVAL data flow system~\cite{NARVAL}. Multi-detector synchronization 
is achieved using the GTS  (Global Trigger and Synchronization System)common clock system~\cite{GTS}, allowing PISTA to couple with {VAMOS++} and EXOGAM (NUMEXO2 digitizers~\cite{Houarner_2025}).

A TGV (Trigger Générique VME) module in the PISTA crate receives validated triggers from the MVLC and requests a $48$-bit, $10$~ns resolution timestamp from the GTS BEAST module. This timestamp is appended to the data of each Mesytec module for every event. 
More details on the implementation can be found in Ref.~\cite{Frankland2025}.

Data are initially read via mvme software~\cite{mvme} and published over the network using ZeroMQ networking library. A software transceiver converts 
these data into the MFM (MultiFrame Metaformat), embedding the timestamp in the header. These frames are then injected into NARVAL data flow, 
where the MFMMerger~\cite{MFMMerger} builds global events using a $1~\mu$s coincidence window.

\section{Characterization of the intrinsic energy detector resolution using radioactive sources\label{sec:res}}
\begin{figure}[t!]
\centering
\includegraphics[width=\columnwidth]{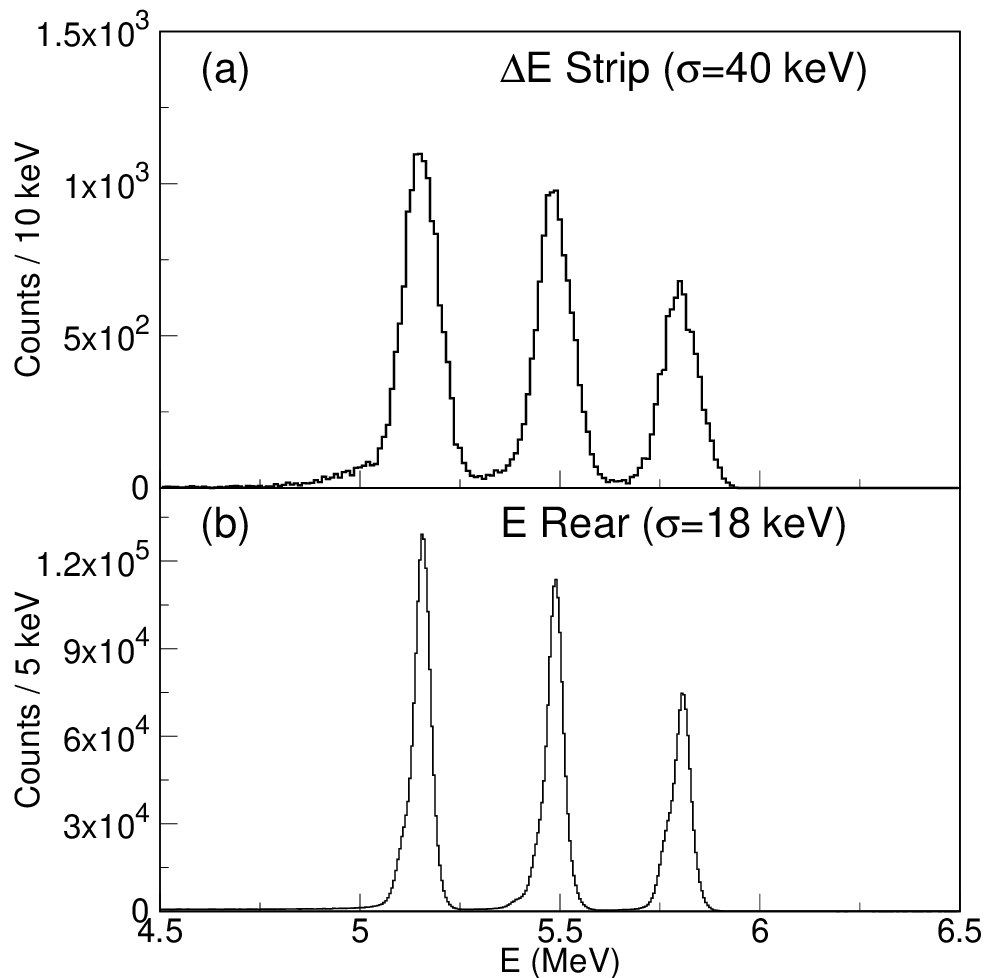}
\caption{\label{pic:Alpha} 
Energy spectra measured using a $3 \alpha$ source are presented. The spectra obtained for (a) a representative strip of $\Delta E$ detector and (b) one representative rear side of $E$ detector.}
\end{figure}

Detectors were characterized using $3\alpha$ calibration sources. 
For the $\Delta E$ detectors, energy resolution was measured to be approximately 
$\sigma\sim40$~keV and $\sigma\sim80$~keV at $\sim5.5$~MeV for the strips and the rear side, respectively. The difference can be attributed to  the substantial capacitance of the rear side. For the $E$ detector, energy resolution was measured to be approximately $\sigma\sim40$~keV 
(limited by the resolution of the MMR ADC for the $200$~MeV range) and $\sigma\sim18$~keV at $\sim5.5$~MeV for the strips and the rear detector, respectively. Figures~\ref{pic:Alpha}(a) and (b) illustrate the typical energy spectra for the $\Delta E$ strip detector and the $E$ rear side detector, respectively, employed for the measurement of the total energy.

\section{Performances using reactions in inverse kinematics}
The performances of the PISTA detector under realistic conditions were determined during an experiment conducted at the GANIL facility in Caen, France. 
The data presented herein were derived from the E850 experimental dataset~\cite{DataE850}.

The PISTA array was positioned in front of the {VAMOS++} spectrometer at a distance of $105$~mm from the $100~\mu$g/cm$^2$ natural Carbon target. 
The target was bombarded by a $5.95$~MeV/A $^{238}$U beam with an intensity ranging between $0.1$ and $2$~pnA. Fission was induced by multi-nucleon transfer reactions in inverse kinematics. The PISTA array was placed around the target at forward angles, symmetrically aligned around the beam axis, 
to detect target-like residues.

Whenever the transfer reaction led to fission, the fragments were transmitted through the central aperture of PISTA to the {VAMOS++} spectrometer, which was positioned at an angle of $20^\circ$. 
The DPS-MWPC detection system, located at the entrance of {VAMOS++}, enabled the event-by-event determination of the trajectory of fission fragments. This capability also allows the determination of the interaction position of the beam on the target.

\subsection{Target-like recoil identification}
\begin{figure}[t!]
\centering
\includegraphics[width=\columnwidth]{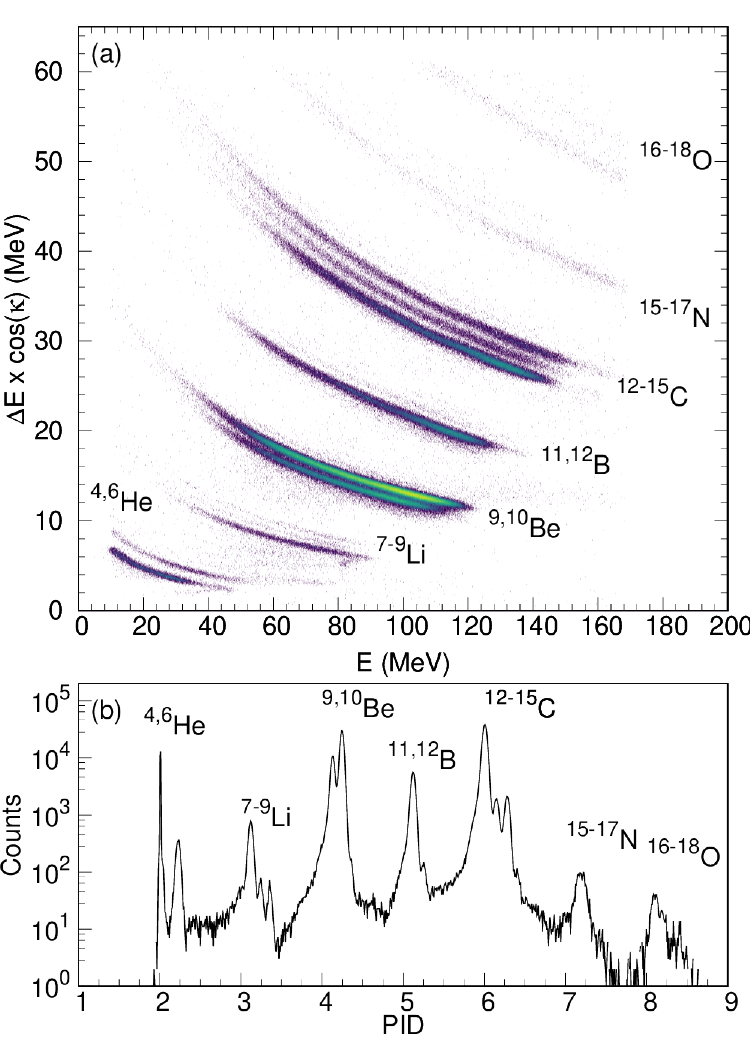}
\caption{\label{pic:dEE} 
Particle identification obtained by PISTA array. 
In panel (a) the correlation between energy loss ($\Delta$E $\times cos(\kappa)$) and residual energy ($E)$ 
for events where the coincident fission fragments were detected in {VAMOS++}, is shown.
In panel (b) the corresponding one-dimensional particle identification (PID) obtained from Deep Neural Network (see text).  
Different detected target-like isotopes are labeled. 
}
\end{figure}

In nuclear reactions occurring at energies close to the Coulomb barrier, multiple reaction channels become accessible. Consequently, unambiguous 
particle identification of the target-like recoil is essential to uniquely identify the fissioning system produced on an event-by-event basis. 
This objective was accomplished using the $\Delta E-E$ method, which necessitates particle penetration until the second stage of the telescope. 
Thus, the thickness of the $\Delta E$ detector determines the identifiable range of ions, encompassing helium to oxygen. 

The $\Delta E$ measurement was derived from the entrance-strip energy signals of the first stage, while the $E$ measurement is obtained from 
the rear-side signals of the second stage. The energy loss ($\Delta E$) was corrected for the difference of effective thickness of the detector 
($cos(\kappa)$ where $\kappa$ denotes the angle between the normal of the detection plane and the particle trajectory). The figure~\ref{pic:dEE} illustrates 
the broad range of populated reaction channels.  The resulting two-dimensional $\Delta E-E$ correlation is presented in Figure~\ref{pic:dEE}(a) 
for three telescopes (with equivalent $\Delta E$ thickness of $103~\mu$m). Isotopic identification is achieved with high precision for ions up 
to oxygen. It should be noted that the nitrogen and oxygen isotopes originate mostly from reactions with contaminants.

A substantial enhancement in selectivity is evident when comparing the results obtained using the PISTA  (Figure~\ref{pic:dEE}(a)) to 
the SPIDER (Figure~4 of Ref.\cite{rodriguez2014}) 
setups. To quantify the improvement, the one-dimensional particle identification (PID) was extracted from the $\Delta E-E$ correlation using a 
Deep Neural Network (DNN) trained on experimental data, following the procedure outlined in Ref.~\cite{Rejmund2025b}.  
The resulting spectrum of PID is shown in Figure~\ref{pic:dEE}(b).
A typical mass resolution of $1.1$~\% (FWHM) for carbon isotopes was obtained, to be compared to $8$~\% reported for SPIDER array. 
Utilizing these resolutions, comparison with Monte Carlo simulations indicates a pixel-level thickness non-uniformity of $\sim0.8~\%$.

\subsection{Target-like recoil kinematics}

\begin{figure}[t!]
\centering
\includegraphics[width=\columnwidth]{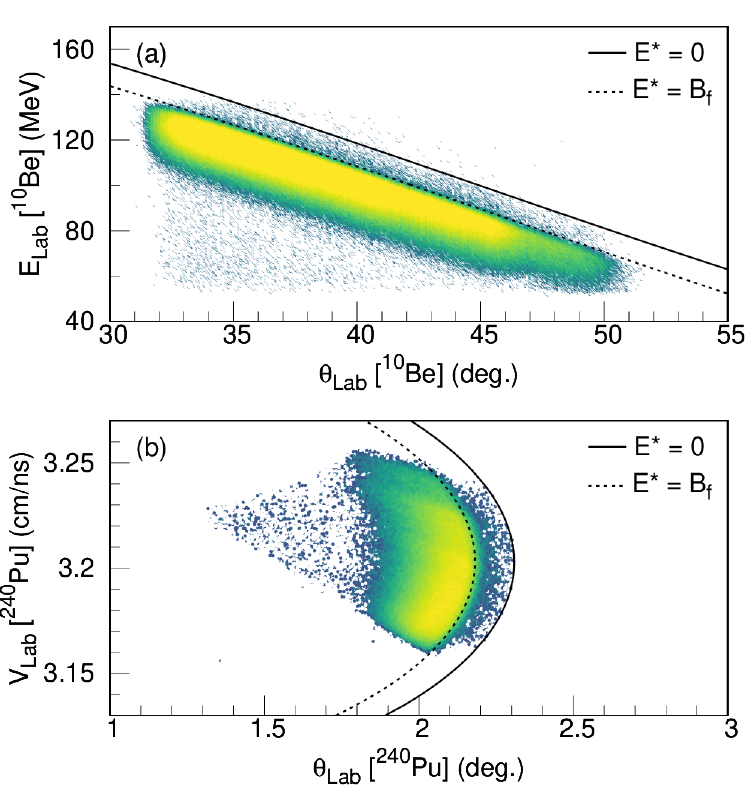}
\caption{\label{pic:kinematics}
Kinematical correlations in the laboratory frame, measured for the $^{12}$C($^{238}$U,$^{240}$Pu)$^{10}$Be reaction.
Panel (a) illustrates the energy correlation with the angle of the target-like recoil $^{10}$Be detected in PISTA, while 
panel (b) illustrates the inferred correlation between the speed  and angle of the $^{240}$Pu fissioning system. 
The calculated kinematic lines for E$^{\ast}=0$ (full line) and E$^{\ast}$=B$_{f}$ (dashed line) are also shown.
}
\end{figure}

The total energy of the light recoils was derived from the sum of $\Delta E$ (obtained from the front side strip signal) and $E$ (obtained from the rear side signal). The $\Delta E$ strip signals were calibrated using a $3 \alpha$ source, supplemented by well-defined elastically scattered $^{12}$C events. The calibration process incorporated kinematic considerations, specifically utilizing the scattering angles derived from the $E$ and $\Delta E$ detector strips. Energy loss calculations were performed for the target, as well as the dead and active layers of the silicon detectors using energy loss tables~\cite{HUBERT1990}. These calculations were essential for determining the energy of the elastically scattered $^{12}$C events used in the calibration. The raw detector signals were correlated with simulated deposited energies and fitted using linear calibration functions.
Following calibration, the deposited energies in the $\Delta E$ and $E$ stages were summed. The kinetic energy at the interaction point was then reconstructed by accounting for the calculated energy losses in both the target material and the detector dead layers. In the data analysis, only events with a single strip multiplicity were considered. The scattering angles of the target-like recoils were inferred from the $\Delta E$ and $E$ strip 
position and the position of the interaction of the beam on the target. The information on the beam interaction position was obtained on an event-by-event basis when reactions lead to fission fragment detected in {VAMOS++}. 
For the events where no fission fragment was detected in {VAMOS++}, the average position of the beam (over the last $1000$ events) was employed to correct for possible beam movements (typical beam spot size was $0.6 \times 1$~mm$^2$). 
A correlation of the energy and scattering angle for the target-like recoil $^{10}$Be measured in PISTA
is shown in Figure~\ref{pic:kinematics}(a) for the $^{12}$C($^{238}$U,$^{240}$Pu)$^{10}$Be reaction. 
Figure~\ref{pic:kinematics}(b) also shows the speed of the $^{240}$Pu fissioning system as a function of its angle derived from the energy and the angle of the target like residue ($^{10}$Be) using energy momentum conservation.

It is noteworthy that the beam-like heavy reaction residue was not detected in {VAMOS++}, and the corresponding properties 
were systematically derived from the properties of the target-like recoil detected in PISTA.
The calculated kinematic lines for $E^* =0$ and $E^{\ast}=B_{f}$, where $B_f$ is the fission barrier, are also shown in Figure~\ref{pic:kinematics}.

\subsection{Reconstruction of the fissioning system excitation energy}

The excitation energy of the fissioning system was determined using the missing-mass method~\cite{Suzuki2012}, which employed the measurement of energy and angle of 
the target-like recoil. This approach assumes that the target-like nucleus is in its ground state and that all excitation energy is imparted to the fissioning system.

\begin{figure}[!t]
\centering
\includegraphics[width=\columnwidth]{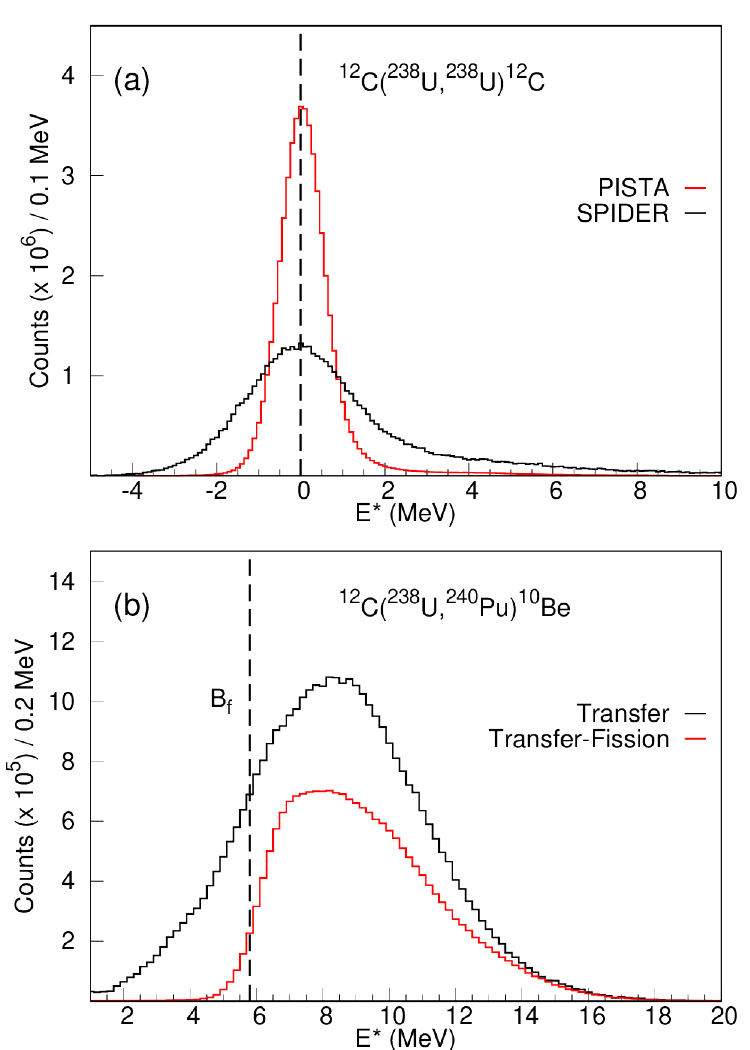}
\caption{\label{pic:Ex}
Reconstructed excitation energy spectra.
Panel (a) illustrates a comparison of the excitation energy spectra in the $^{12}$C($^{238}$U,$^{238}$U)$^{12}$C elastic channel 
obtained with PISTA array (red line) and SPIDER array (black line)~\cite{rodriguez2014}. 
Panel (b) shows the excitation energy spectra for the $^{12}$C($^{238}$U,$^{240}$Pu)$^{10}$Be reaction channel 
for inclusive transfer reaction events (black line) and transfer reaction in coincidence with fission events detected in {VAMOS++} (red line). 
See text for further details. The energy of the fission barrier ($B_f = 5.6$~MeV~\cite{Bjornholm1980}) is indicated by a dashed line.
}
\end{figure}

Figures~\ref{pic:Ex} (a) and (b) depict the excitation energy spectra for reactions resulting in $^{12}$C and $^{10}$Be target-like recoils, respectively. 
The excitation energy resolution can be assessed using the elastic channel ($^{12}$C($^{238}$U,$^{238}$U)$^{12}$C with $E^{\ast}$=0). 
A resolution of $\sigma_{E} =518(1)$~keV was determined by a gaussian fit on the left side of the elastic distribution to exclude the possible contribution from inelastic events. This value can be compared to $\sigma_{E}=1150(10)$~keV 
reported in the case of SPIDER array~\cite{rodriguez2014}. It is important to note that in the elastic channel $^{238}$U nuclei do not undergo fission, rendering it impossible to determine the beam interaction position on the target on an event-by-event basis.  The average beam interaction position was employed to account only for average beam displacement, while the beam spread remained unaccounted for. Consequently, the newly reported resolution represents an upper limit for fission events.

Figure~\ref{pic:Ex} (b) presents the reconstructed excitation energy spectra obtained for the $^{12}$C($^{238}$U,$^{240}$Pu)$^{10}$Be channel. 
The red histogram shows the excitation energy distribution of $^{240}$Pu detected in PISTA in coincidence with a fission fragment in {VAMOS++} 
corrected for the acceptance of the {VAMOS++} spectrometer 
$$N_{Fis} = N_{V}/a,$$ 
where: $N_{V}$ is the number of $^{10}$Be nuclei detected in PISTA in coincidence with a fission fragment in {VAMOS++}
and $a$ is the simulated acceptance of {VAMOS++}. The acceptance of VAMOS was simulated to be $0.065$,  utilizing the fission fragments distribution of $^{240}$Pu calculated by GEF~\cite{GEF}. The realistic excitation energy distribution of the fissioning system was employed. The black histogram shows the inclusive distribution in excitation energy for transfer events 
$$N_{Tr} = N_{Fis}  +  N_{\bar{V}} \times D,$$ 
where: 
$N_{\bar{V}}$ is the number of $^{10}$Be nuclei detected in PISTA with a downscaling factor $D=300$ for
the PISTA experimental trigger, for which no fragments are detected in VAMOS. 
It can be seen from the Figure~\ref{pic:Ex}(b), that in this case, the 2p-transfer channel populates states in $^{240}$Pu up to approximately 
$16$~MeV of excitation energy.  The rapid increase of the fission probability obtained in coincidence with fission fragments in the vicinity of 
the fission barrier~($B_f$)~\cite{Bjornholm1980}, indicated by the dashed line, can be utilized to infer excitation energy resolution.

Monte Carlo simulations were conducted to assess the various factors contributing to the obtained excitation energy resolution. In the case of elastic scattering, the resolution in excitation energy was decomposed into several components. The target thickness contributed an uncertainty of $\sigma~\sim~208$~keV, while the target-like recoil interaction position within the PISTA array introduced an additional uncertainty of $\sigma~\sim~124$~keV. The intrinsic energy resolution resulted in an uncertainty of $\sigma~\sim~126$~keV. An uncertainty in the beam spot size of $\pm 50\mu$m was used to evaluate the beam size contribution ($\sigma_x = 350 \pm 50$~$\mu$m and $\sigma_y = 550\pm50$~$\mu$m). This leads to the largest contribution to the resolution in excitation energy of $\sigma~\sim~419~\pm~46$~keV. It should be noted that this contribution varies strongly from one telescope to another due to the asymmetry of the beam spot size between vertical and horizontal directions. The combined effect of these factors yielded a total average uncertainty in excitation energy of $\sigma_E~\sim~500~\pm~39 $~keV. This mean value slightly underestimates the experimental data, but reveals the large sensitivity to uncertainties in the beam size (measured using DPS-MWPC with an inherent uncertainty).

In the case of the $^{12}$C($^{238}$U,$^{240}$Pu)$^{10}$Be reaction channel, similar simulations were conducted using realistic excitation energy distributions. The uncertainty in the energy of the reaction, attributed to the target thickness, accounts for $\sigma~\sim~175$~keV. The uncertainty in the target-like recoil interaction position within the PISTA array is responsible for $\sigma~\sim~85$~keV. The influence of the intrinsic energy resolution contributes to $\sigma~\sim~75$~keV. The differences compared to the $^{12}$C case in these contributions can be understood given the difference in the kinematics of the reactions. Regarding the contribution of the beam spot, in the present case, events detected in coincidence with a fission fragment allowed for an accurate determination of the beam interaction position on the target on an event-by-event basis. Using a position resolution on the target of $\sigma~\sim~400$~$\mu$m results in a contribution of $\sim~270$~keV to the excitation energy resolution. It should be noted that, in contrast to the case of $^{12}$C, the uncertainty in the intrinsic beam spot size does not affect the resolution and that contribution is dominated by the resolution on the determination of position on target.  Consequently, the total resolution in excitation energy was determined to be $\sigma_E~\sim~340$~keV for events in coincidence with a fission fragment.

\subsection{Target-like recoil coincidence with $\gamma$ rays}
\begin{figure}[t!]
\includegraphics[width=\columnwidth]{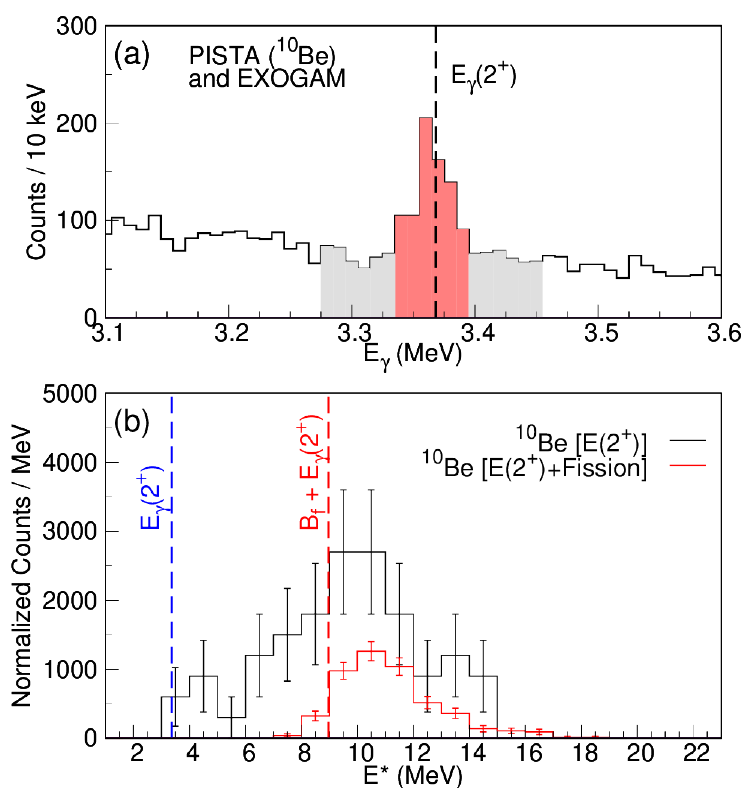}
\caption{\label{pic:gamma}
Coincidences with $\gamma$ rays.
Panel (a) illustrates the Doppler-corrected- $\gamma$-ray spectra measured by EXOGAM in coincidence with $^{10}$Be detected in PISTA. 
The energy of the  first excited state, $E_\gamma(2^+)=3.368$~MeV,  is indicated  by a dashed line. 
The red and gray shaded area illustrate respectively the regions used for selection of peak and background events used in background subraction proceedure. 
Panel (b) depicts the distribution of excitation energy inferred from $^{10}$Be measured in coincidence with the $E_\gamma (2^+)$ 
$\gamma$-ray transition of $^{10}$Be (black) and with additional coincidence with fission in {VAMOS++} (red).  
The blue and red dashed lines, respectively, indicate the energy $E_\gamma (2^+)$ and $B_f + E_\gamma (2^+) $.
}
\end{figure}

The PISTA array’s mechanics was designed to facilitate its coupling to other detectors and in particular with the HPGe clover detectors of EXOGAM~\cite{EXOGAM}.In the present experiment, three EXOGAM clovers were strategically positioned at backward angles of approximately $135^\circ$, situated at a distance of 
$120$~mm from the target. $\gamma$-ray addback technique was employed to ensure the recovery of complete energy 
deposits within the EXOGAM clover detectors. The highest energy deposit within each segment served as the basis for determining 
the $\gamma$-ray detection angle. The measured $\gamma$-ray energy was corrected for Doppler effects event-by-event using the velocity vector measured with PISTA, in order to obtain the corresponding $\gamma$-ray spectrum emitted by the target-like recoil.

The reconstruction of the excitation energy is based on the hypothesis that the entire excitation energy is carried by the beam-like fragment. To assess the validity of this assumption, the detection of $\gamma$-rays in coincidence with target-like nuclei in the PISTA array can be utilized to estimate the probability of exciting the target-like nucleus.  Figure~\ref{pic:gamma}(a) depicts the Doppler-corrected $\gamma$-ray spectrum of events in coincidence with $^{10}$Be.
The first excited state of $^{10}$Be, characterized by an energy of $E_\gamma (2^+) = 3.368$~MeV, is evident in the spectrum, 
thereby demonstrating that $^{10}$Be can be populated in its excited states in the $^{12}$C($^{238}$U,$^{240}$Pu)$^{10}$Be reaction; which is consistent with the results of Ref.~\cite{rodriguez2014}. 
Figure~\ref{pic:gamma}(b) depicts the distribution of excitation energy in $^{240}$Pu, assuming $^{10}$Be was populated in its ground 
state, measured in coincidence with the $3.368$~MeV transition.

The spectrum was obtained using the $\gamma$-ray gating region highlighted in Fig.~\ref{pic:gamma}(a) by the red shaded area for the peak and the gray shaded area for the background. Background events were scaled to the peak region and subtracted from the peak events.  In Fig.~\ref{pic:gamma}(b), the black and red lines represent, respectively, the distributions of excitation energy measured for transfer and transfer-induced fission events (in coincidence with fission fragments detected in VAMOS++). The $\gamma$-ray gated distribution of [$^{10}$Be and fission] and [$^{10}$Be] were respectively normalized to account for the acceptance of VAMOS and the trigger reduction factor $D=300$. A threshold for $\gamma$-ray emission is evident in the figure, red line, at $E^{\ast} = B_f + E_\gamma (2^+) = 8.94$~MeV. For energy levels below  this threshold, when the target-like $^{10}$Be nucleus undergoes excitation, the excitation energy of the complementary  $^{240}$Pu system  is insufficient to overcome the fission barrier. A similar behavior can be observed from the inclusive transfer distribution, black line,  with the threshold at $E^{\ast}=E_\gamma (2^+)$ (shown with blue dashed lines).  This measurement of $\gamma$-rays in coincidence with $^{10}$Be enables the determination of the probability of exciting target-like nuclei. In this specific case, the probability was measured to be $10~(3)$~\%. This probability can subsequently be accounted for when extracting fission probabilities~\cite{rodriguez2014}.

\section{Conclusions and outlook}

In the present work, we report on a new detector array, PISTA, designed to characterize fissioning systems populated in multi-nucleon transfer 
reactions in inverse kinematics. We focus on the measurements providing simultaneously, on an event-by-event basis, the high-resolution excitation energy of the fissioning system and the complete isotopic identification of the fission fragments.  The geometry of the PISTA array enables isotopic identification of the target-like nuclei up to oxygen isotopes and with a mass resolution of $1.1$~\% (FWHM).  An upper limit of $\sigma_E = 518~(1)$~keV, corresponding to $FWHM_E=1.217~(2)$~MeV, was determined for the excitation energy resolution in the elastic channel when the beam position on the target could  not be precisely measured. For events where a fission fragment was detected by the {VAMOS++} spectrometer, the target position could be reconstructed. 
In this case, the corresponding Monte Carlo simulations yielded an excitation energy resolution of $\sigma_E \sim 340$~keV, corresponing to $FWHM_E \sim 800$~keV. The present results represent a significant improvement compared to previous experimental setups, which will enable, for a wide range of fissioning systems in the actinide region, the extraction of fission probabilities and isotopic fission yields as a function of the excitation energy of the fissioning nucleus, measured with a better precision than with previous setup. In the future, the high spatial segmentation of PISTA, will also enhance the selectivity of event reconstruction of reaction involving multiparticle events like in the reaction channel $^{12}$C($^{238}$U, $^{242}$Pu)$^{8}$Be, where the $^{8}$Be is unbound and decays through the emission of two $\alpha$ particles. While similar study was already performed with SPIDER in the $^{9}$Be($^{238}$U, $^{239}$U)$^{8}$Be reaction~\cite{RamosPRL2019}, the advantages of PISTA are twofold: (i) the high granularity minimizes the pile-up of the two $\alpha$ 
particles in the same strips, and (ii) the improved angular resolution results in an enhanced excitation energy resolution. This will allow the efficient detection of both $\alpha$ particles and the reconstruction of the excitation energy with unprecedent precision. 

\section*{Acknowledgments}
We acknowledge the important technical contributions of
J. Goupil, L. Menager and A. Giret and the GANIL accelerator staff.
L. B.-G. and A.F. acknowledge support from the Région Normandie for the grants supplied under the Réseaux d'Intérêts Normands (RIN) Doctorants 50\% and Emergents SASAV, respectively. This work was partially supported by the Spanish Research State Agency nº PID2021-128487NB-I00, by European Union ERDF, by the "María de Maeztu” Units of Excellence program nº MDM-2016-0692, by Xunta de Galicia as Centro singular de investigación de Galicia accreditation 2019–2022, and from the "Consolidación e estruturación” nº ED431F 2016/002. This work was performed under the auspices of the U.S. Department of Energy by Lawrence Livermore National Laboratory under Contract DE-AC52-07NA27344, and supported in part by the Laboratory Directed Research and Development (LDRD) Program. The views expressed in this article are those of the authors and do not reflect the official policy or position of the U.S. Naval Academy, Department of the Navy, the Department of Defense, or the U.S. Government.

\bibliographystyle{myapsrev4-1}
\bibliography{references}

\end{document}